\documentclass[runningheads]{svmult}

\usepackage{makeidx}   
\usepackage{graphicx}  
\usepackage{subeqnar}  
\usepackage{multicol}  
\usepackage{physprbb}  
\makeindex             

\newcommand{\Xs}{X$_{\odot}$} 
\newcommand{\Xm}{X$_{ISM}$}
\newcommand{\lis}{$^7$Li/$^6$Li }
\newcommand{\bb}{$^{11}$B/$^{10}$B }

\begin{document}
\title*{Li, Be, B and Cosmic Rays in the Galaxy}
\toctitle{Evolution of light elements}
%
%
\titlerunning{Li, Be, B and Cosmic Rays in the Galaxy}
%
\author{Nikos Prantzos}
\authorrunning{N. Prantzos}
%
%
\institute{Institut d'Astrophysique de Paris, 98bis Bd Arago, Paris, France}

\maketitle              

\begin{abstract}
A short overview is presented of current issues concerning the production
and evolution of Li, Be and B in the Milky Way. It is argued that the
currently popular idea that Galactic Cosmic rays are accelerated inside 
metal-rich superbubbles (which leads ``naturally'' to the production of 
primary Be and B, as observed) encounters the same problems as the 
previously popular idea of supernovae accelerating their own ejecta.
A major challenge to theories of light element production is presented
by the recent (and still preliminary) data suggesting a 
surprisingly high and $\sim$constant abundance of $^6$Li in
halo stars; attempts to explain such a ``plateau'' 
are critically examined.
\end{abstract}

\section{The 1970s: problems with late $^7$Li, $^{11}$B}
The idea that the light and fragile elements Li, Be and B are
produced by the interaction  of the energetic nuclei
of galactic cosmic rays (CGR) with the nuclei of the interstellar medium
(ISM) was introduced 35 years ago (Reeves et al. 1970, Meneguzzi et al. 1971).
In those early works it was shown that, taking into account
the relevant cross-sections and with plausible assumptions about the
GCR properties (injected and propagated spectra, intensities etc.; see
Fig. 1, {\it right column})
one may reproduce reasonably well the abundances of those light
elements observed in meteorites {\it and }   in GCR.

Two problems were identified with the GCR production, compared
to meteoritic composition: 
the \lis ratio ($\sim$2 in GCR but $\sim$12 in meteorites) and the \bb
ratio ($\sim$2.5 in GCR but $\sim$4 in meteorites). Modern solutions to
those problems involve {\it stellar} production of $\sim$70\% of
$^7$Li (in the hot envelopes of AGB stars and/or novae) and of $\sim$40\% 
of $^{11}$B (through $\nu$-induced spallation of $^{12}$C in SNII). In both
cases, however, uncertainties in the yields are such that observations
are used to  constrain the yields of the candidate sources
rather than to confirm the validity of the scenario.

\section{The 1980s: the Li plateau; primordial, but low or high?}

One of the major cosmological developments of the 1980s was the
discovery  of the Li plateau in low metallicity halo stars
(Spite and Spite 1982, see Fig. 1 {\it top left} panel).
The unique behaviour of that element, i.e. the constancy of
the Li/H ratio with metallicity, strongly suggests a primordial
origin. The observed value has been extensively used (along with
those of D and $^4$He) to constrain the physics of primordial
nucleosynthesis and, in particular, the baryonic density
of the Universe (e;g. Steigman, this meeting). In particular, the difference
between the observed plateau value (Ryan et al. 1999) and the Li
abundance corresponding to the baryonic density derived from
WMAP data (Fig. 1) is rather high ($\sim$0.5 dex) and points
to a failure of our understanding, of either stellar atmospheres, 
primordial nucleosynthesis
or Li depletion in stars (e.g. Lambert 2004 and references therein).


\begin{figure}[t!]
\begin{center}
\includegraphics[angle=-90,width=\textwidth]{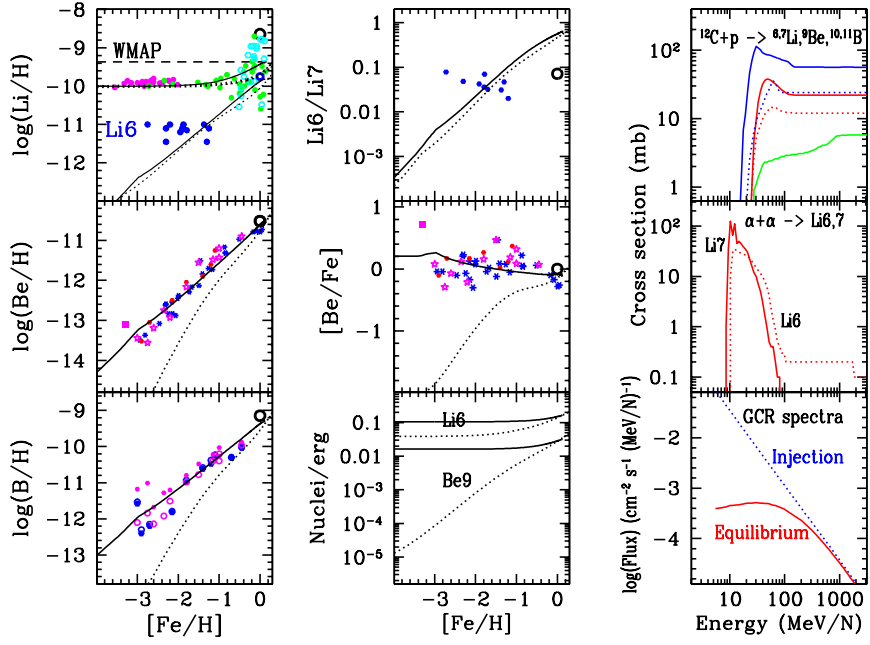}
\end{center}
\caption[]{{\it Left and middle columns}: Evolution of LiBeB 
with two different compositions assumed for GCR
(1) X$_{GCR}(t)$ = f$_{ENH}$ \Xm(t) always ({\it dotted curves}) and (2) 
 X$_{GCR}$ = f$_{ENH}$ \Xs \ always, ({\it solid curve}), i.e. 
metallicity dependent
and independent compositions, respectively (X representing {\it mass 
fraction}); 
in both cases, the {\it enhancement factor} 
f$_{ENH}$ is such as to match the present day observed GCR composition, 
which is
enriched in C (by $\sim$8), N  (by $\sim$2) and O (by $\sim$5), i.e. the
``metallicity'' of the GCR  fluid today is about 5 times solar in the solar 
vicinity.
In both cases, the GCR injection spectrum is Q(E,t) = S \ q(E) R$_{SN}$(t),
where  R$_{SN}$(t) is the SN rate,
q(E) is displayed in  {\it right bottom panel}
and the normalisation factor S is such that
the energy of all nuclei in GCR is
$\sum_i X_{i,GCR} \int q(E) E dE$=$\epsilon_{CR}$E$_{KIN}$ 
(where the kinetic energy E$_{KIN}$=1.5 10$^{51}$ ergs per 
SN and $\epsilon_{CR}\sim$0.1-0.2
is a typical acceleration efficiency of SN). Model (2) produces naturally
primary Be and B with reasonable (and metallicity $\sim$independent)
values of  $\epsilon_{CR}$, while Model (1) requires metallicity dependent 
values
which become unacceptably large at low metallicities ({\it middle bottom} 
panel).
Despite its success with Be and B, Model (2) cannot produce the Li6 plateau.
The evolution of C,N,O,Fe is followed 
with metallicity dependent yields, but no Li7 from Hot-bottom burning in 
AGB stars or B11 from $\nu$-nucleosynthesis in massive stars is included.
As a result, the abundances of Li7, B11 and Li6/Li7 at [Fe/H]=0 differ from
their solar values (by factors 6, 2 and 6, respectively).
Production cross-sections from $^{12}$C+p (for all
light isotopes) and from $\alpha$+$\alpha$ (for Li6, Li7), as
well as GCR spectra at the source (injection) and
after propagation 
(equilibrium) are displayed in the {\it right column}.
The {\it equilibrium spectrum} should be folded with the corresponding 
cross-sections and abundances to
calculate relevant production rates of LiBeB. Energies are in MeV/nucleon.
}
\label{eps1}
\end{figure}

\section{The 1990s: problems with early Be and B (primaries!)}
Observations of halo stars in the 90s revealed a linear relationship
between Be/H (as well as B/H) and Fe/H. That was unexpected, since Be and B
were thought to be produced as {\it secondaries}, by spallation of the 
increasingly abundant CNO nuclei. Only the Li isotopes, produced at low
metallicities mostly by $\alpha+\alpha$ fusion reactions (Steigman and Walker
1992) are produced as primaries. The only way to produce primary Be and B is
by assuming that GCR have always the same CNO content (Duncan et al. 1992). 
The most convincing argument in that respect is the "energetics
 argument"
put forward by Ramaty et al. (1997): if SN are the main source of GCR energy,
there is a limit to the amount of light elements produced per SN, which depends
on GCR and ISM composition. If the metal content of both ISM and GCR becomes
low, there is simply not enough energy
in GCR to keep the Be and B yields constant (as required by observations).
Since the ISM metallicity certainly increases with time, the only possibility
to have $\sim$constant LiBeB yields is by assuming that GCR have a
$\sim$constant metallicity.

A $\sim$constant abundance of C and O in GCR 
can naturally be understood if SN
accelerate their own ejecta. However, the absence of unstable $^{59}$Ni 
(decaying through e$^-$-capture
within 10$^5$ yr) from observed GCR suggests that acceleration occurs 
$>$10$^5$ yr after the explosion (Wiedenbeck et al. 1999) 
when SN ejecta are presumably 
diluted in the ISM.  Higdon
et al. (1998) suggested  that GCR are accelerated in {\it superbubbles} 
(SB), enriched by the ejecta of many SN as to have a large 
and $\sim$constant metallicity. Since then, this became {\it by default}, 
the  "standard" scenario for the production of primary Be and B by GCR,
invoked in almost every  work on that topic.

However, the SB scenario suffers from (at least) two problems.
First,  core collapse SN are observationally associated
to HII regions (van Dyk et al. 1996) and it is well known that the
metallicity of HII regions reflects the one of the {\it ambient ISM} (i.e.
it can be very low, as in IZw18) rather than the one of SN. 
Moreover, Higdon et al. (1998) evaluated the time interval $\Delta t$ between
SN explosions in a SB to a 
comfortable $\Delta t \sim$3 10$^5$ yr, leaving enough time to $^{59}$Ni 
to decay before the next SN explosion and subsequent acceleration. 
However,  SB are constantly powered not only by SN
but also by the strong  {\it winds of
massive stars} (with integrated energy and acceleration
efficiency similar to the SN one, e.g. Parizot et al. 2004), 
which should  continuously
accelerate $^{59}$Ni, as soon as it is ejected from SN explosions. 
Thus, SB suffer exactly from the same problem that plagued SN as accelerators
of metal rich ejecta.

The problem of the acceleration site of GCR (so crucial for the
observed linearity of Be and B vs Fe) has not found a satisfactory 
explanation yet.

\begin{figure}[t]
\begin{center}
\includegraphics[angle=-90,width=0.9\textwidth]{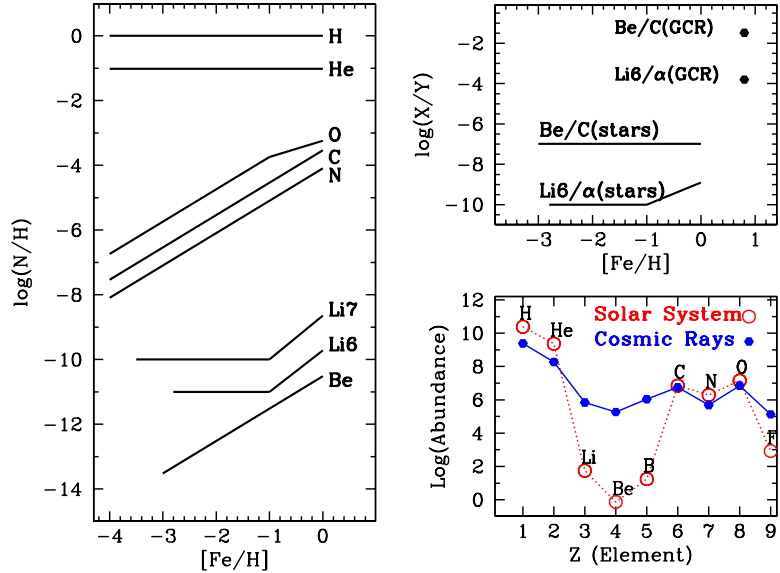}
\end{center}
\caption[]{{\it Left:} Schematic evolution of the light elements,
from H to O, in the local Galaxy, according to observations 
(the evolution of B, similar to the one of Be, is 
not shown for clarity). {\it Right top:} Schematic
evolution of abundance ratios
Be/C  and Li6/He (by number) in the local Galaxy. Those ratios are chosen
because Be is produced mainly by C (and N,O, which display similar
evolution) while Li6 is
produced mainly by $\alpha+\alpha$ (especially in the early Galaxy,
while production by CNO becomes important later). In  both cases,
the constancy of those ratios is reminiscent of an ``equilibrium'' process
(with the production rate balanced by the destruction rate, as e.g. in
the CNO cycle operating at equilibrium). In GCR, the light isotopes
are certainly at equilibrium with their ``father'' nuclei,
since the observed abundances ({\it Right bottom}) correspond indeed
to such equilibrium values (ratios displayed at {\it Right top}); 
moreover, the ratio of (X/Y)$_{GCR}$/(X/Y)$_{ISM}$
is the same for Be/C and Li6/He (around 10$^5$-10$^6$). However, although  it
is easy to understand equilibrium abundances of light elements in GCR
(where production and destruction by the abundant H and He
of the ISM medium are rapid), it is difficult to conceive such
an equilibrium situation for the nuclei residing in the ISM
(where production and destruction by the rarefied H and He
gas of the GCR  are very slow).
}
\label{eps1}
\end{figure}

\begin{table*}
\caption{Problems with LiBeB and GCR
\label{fig:Table2}}
\begin {center}
\begin{tabular}{ccc}
\hline \hline
{\bf Problem} & {\bf Suggested solution} & {\bf Comments}  \\
\hline
{\bf Late Li: $^7$Li/$^6$Li}   & Late $^7$Li (but not $^6$Li)      & Plausible, but Li yields of AGB    \\
Solar value=12 & from AGB and novae   & and rates/yields of novae   \\
but GCR=2    &       &   VERY uncertain          \\
     &       &             \\
{\bf Late B: $^{11}$B/$^{10}$B }   &  40\% of $^{11}$B  produced     & Plausible, although   \\
Solar value=4 & by $\nu$-nucleosynthesis   &  $\nu$-spectra are uncertain \\
but GCR=2.5    &  in SNII     & Produces {\it primary}    $^{11}$B          \\
     &       &             \\
{\bf Early Be and B}   &  GCR metallicity always   & Problem with absence of unstable  \\
Observations: {\it primaries}, & the same, originating in  &  $^{59}$Ni in GCR; it becomes stable  \\
while ``Standard'' GCR & SN ejecta or in & if directly accelerated in SN \\
produce {\it secondaries}     & {\it Superbubbles} (SB)      &  or continuously accelerated     \\
     &       &   in SB by {\it stellar winds}       \\
     &       &             \\
     & 1){\it  Primordial}, non-standard      &  Particles and cross-sections           \\
     &  production in BBN    &   unknown          \\ 
 {\bf Early $^6$Li/H :} &  2) {\it Pregalactic} production   & CR energetics unknown;   \\
  A too high        &   during structure formation    & hard to explain ``plateau'' \\ 
   ``plateau''   & 3)  {\it Equilibrium}: $^6$Li/$\alpha\sim$const. &  Requires too rapid reactions,  \\
      &   Production=Destruction    & incompatible with GCR densities            \\
\hline \hline
\end{tabular}
\end{center}
\end{table*}

\section{The 2000s: a $^6$Li plateau? primordial or (pre-)galactic?}
 
The recent report of a "plateau" for $^6$Li/H in halo stars
(Asplund et al. this meeting) gives a new twist to the 
LiBeB saga. The detected $^6$Li/H (and corresponding \lis) value
at   [Fe/H]=-2.8 is much larger than expected if GCR are the only source
of the observed Be/H and $^6$Li/H in that star (see Fig. 1). 
A few explanations have been proposed for such a high early amount
of  $^6$Li:

1) Primordial, non-standard, production during Big Bang Nucleosynthesis
(BBN): the decay/annihilation of some massive particle (e.g.
neutralino) releases energetic nucleons/photons which produce $^3$He or  $^3$H
by spallation/photodisinte- gration of  $^4$He, while
subsequent fusion reactions between $^4$He and $^3$He or  $^3$H 
create  $^6$Li (e.g. Jedamzik 2004). Observations of  $^6$Li/H constrain
then the masses/cross-sections/densities of the massive particle. 

2) Pre-galactic, by fusion reactions of  $^4$He nuclei, accelerated
by  the energy released during structure formation (Suzuki and Inoue 2002);
in that case, CR energetics are decoupled from SN energetics. 
In view of the many uncertainties related to the behaviour of
the baryonic component during structure formation, the energetics of CR
in that scenario are very poorly known/constrained at 
present\footnote{Fields and Prodanovic (2004) suggest 
that the associated production
of $\gamma$-rays (from decaying pions, produced py energetic $p + p$ reactions)
could contribute significantly to the extragalactic $\gamma$-ray background;
see also Prantzos and Cass\'e (1994).}.
Moreover, in order 
to explain the observed $^6$Li ``plateau'' the effect must end 
(or drastically decrease) {\it before} the first
stars form and explode releasing Fe, otherwise $^6$Li/H would
continue increasing at metallicities [Fe/H]$>$-3. But this 
runs against  our current understanding of
structure formation, which suggests that merging of sub-structures
continues with mildly reduced intensity 
during a large fraction of the Galaxy's early life 
(e.g. Helmi et al. 2003).

3) A third possibility is suggested by the  observed $\sim$constancy
of $^6$Li/He and Be/C with [Fe/H] (Fig. 3 {\it right top}), 
reminiscent of an {\it equilibrium}  process.
Indeed, the abundances of the LiBeB isotopes {\it in GCR} are much
higher than the solar ones ({\it right bottom}) and are in equilibrium
with those of the progenitor He,C,N,O nuclei, 
i.e. the ratio of daughter/progenitor abundances in GCR
is roughly equal to the one of the
corresponding production/destruction spallation cross-sections.
The advantage of that idea is that it explains at one stroke both
the primary Be and B (always at equilibrium with progenitor CNO nuclei)
and the $^6$Li plateau (with  $^6$Li at equilibrium with
progenitor $^4$He).
However, equilibrium requires very fast production and destruction reactions
(with timescales shorter than the evolutionary timescale of the system);
this certainly happens inside  GCR (with the fast GCR particles interacting
with the numerous ISM nuclei) but the opposite does not 
hold\footnote{The interaction timescale is: 
$\tau \sim$(n $\sigma v$)$^{-1}$.
For  GCR $^6$Li interacting with
ISM protons of density $n_p\sim$1 cm$^{-3}$ one has
$\tau \sim$10$^7$ yr, but for ISM  $^6$Li interacting with
GCR protons of $n_p\sim$10$^{-9}$ cm$^{-3}$ 
(corresponding to the observed energy density of 1 eV/cm$^3$)
$\tau$ is a billion times larger ($\sigma \sim$400 mb 
being the destruction cross-section).}.

%

\end{document}